\documentclass[preprintnumbers,nofootinbib,floats,floatfix]{revtex4-2}
\usepackage{amsfonts}
\usepackage{amsmath}
\usepackage{graphicx}
\usepackage{amssymb}
\usepackage{amsxtra}
\usepackage{mathrsfs}  
\usepackage{amstext}
\usepackage{orcidlink}
\usepackage{endnotes}
\usepackage{hyperref}
\hypersetup{colorlinks,linkcolor={cyan},citecolor={cyan},urlcolor={magenta}}
\begin{document}
\title{Constrained Hamiltonian dynamics of 3D gravity coupled to topological matter}
\author{Omar Rodr{\'i}guez-Tzompantzi} \email{omar.tzompantzi@unison.mx}
\affiliation{Departamento de Investigaci\'on en F\'isica, Universidad de Sonora,
Apartado Postal 5-088, C.P. 83000, Hermosillo, Sonora, M\'exico.}
\begin{abstract}
We present the Dirac Hamiltonian formalism for a pair of $1$-form fields with a topological-like potential coupled to first-order gravity in three-dimensional spacetime.  By considering the complete phase space, we derive the full structure of the physical constraints, including both primary and secondary ones; analyze their consistency conditions; classify them into first- and second-class; and compute their Poisson-bracket algebra. Our analysis confirms the absence of local degrees of freedom, consistent with the topological nature of the model’s action. Furthermore, we construct the canonical generator for gauge transformations and demonstrate that, through appropriate gauge parameter mappings, these transformations recover the full diffeomorphism and Poincaré symmetries of the Lagrangian formulation. Finally, we explicitly compute the Dirac brackets, establishing the symplectic structure of the reduced phase space.
\end{abstract}
\preprint{}
\maketitle
\section{Introduction}
\label{introduction}
One of the most challenging and fascinating problems in modern theoretical physics is the formulation of a consistent quantum theory that combines the principles of Einstein's general relativity with quantum mechanics. Such a theory would be crucial not only for addressing the unification of fundamental forces \cite{Isenberg} but also for understanding the nature of gravitational singularities \cite{Penrose}. However, constructing soluble models of quantum gravity is notoriously difficult due to non-renormalizability \cite{Hooft,Stelle,Goroff,Goroff2} and other fundamental obstacles \cite{Polchinski,Polchinski2,Rovelli,Thiemann,Buoninfante,Edward}. In light of these challenges, the study of lower-dimensional gravity models has become highly valuable. A common feature of these models is the absence of local degrees of freedom, a property that renders them significantly more tractable than their higher-dimensional counterparts. In particular, their simplified structure allows for a complete analysis of the classical dynamics, which is critical both for understanding the classical theory and for developing non-perturbative approaches to quantization \cite{Teitelboim,Polchinski3, Alexander,J-T,J-T2,Deser,Deser2}. These models provide a powerful framework for elucidating conceptual problems and exploring quantum-gravitational aspects while keeping technical complexities manageable. Prominent examples, and the central focus of this work, are  gravity theories in three dimensions. 

In three dimensions, pure Einstein gravity, with or without a cosmological constant, lacks local degrees of freedom yet retains the core conceptual structure of the four-dimensional theory. It is fundamentally a topological theory, devoid of gravitational waves or action-at-a-distance phenomena \cite{Deser,Deser2}. A key feature is that it can be recast as a gauge theory, with an action principle constructed from one-form fields—the vielbeins and the dualized spin connections—valued in the Lie algebra of the three-dimensional ($3$D) Lorentz group \cite{Achucarro,Witten2,Witten,Horne}. The action is of the topological type, meaning that no background metric is required in this formalism. Thus, this first-order formulation intrinsically links symmetries, topology, and geometry. Consequently, 3D gravity is a remarkably rich theory; it contains non-trivial solutions such as the BTZ black hole with its thermodynamic behavior \cite{BTZ,Banados,Black-hole,BTZ-EXOTIc}, exhibits profound holographic properties \cite{Holography,Coussaert,Dittrich,Hikida}, and allows for non-perturbative quantization through various approaches \cite{Carlip,Viqar,Noui,Freidel,Porrati,Meusburger,Bonzom}. Moreover, its mathematical framework has become a valuable tool for modeling the mechanisms behind exotic physical phenomena observed in certain condensed matter systems \cite{Vera,Tekin,Smolyaninov,Igor}.

However, to model realistic physical systems or to enrich its classical and quantum structure, coupling matter to 3D gravity is necessary. Introducing matter fields into this context presents unique challenges: the main difficulty stems from the topological nature of pure 3D gravity. It turns out that coupling matter in the traditional way—via the metric—generally destroys this topological character; consequently, the quantization process faces the same obstacles as its $4$D counterpart \cite{Romano,Damiano}. Crucially, however, in the first-order formalism, matter can be coupled directly through the connection. This approach preserves the solvability inherited from pure 3D gravity. In fact, this framework reveals that certain special classes of topological field theories constitute notable exceptions to the aforementioned issue of insolvability in the presence of matter \cite{Romano,Gegenberg1,Burwick,SCarlip,Marcela,Horowitz}. 

In this work, we shall focus on one such exceptional class: the coupling of topological matter fields to first-order gravity in three dimensions. We concentrate specifically on 1-form fields, as they provide the simplest non-trivial matter content while preserving the solvability of 3D pure gravity. Such couplings are known to support gauge-invariant topological mass generation mechanisms, offering a natural framework to study massive excitations in a gravitational context \cite{Afshar,Bergshoeff}. In cosmology, the consideration of $1$-form fields has recently gained relevance as a means to propose inflationary models or to explore a possible electromagnetic origin for the cosmological constant \cite{Ford,Himmetoglu}. This choice is further motivated by their prominent role in condensed matter physics, where they describe fractionalization phenomena—as seen in fractional quantum Hall systems, spin liquids, and other topological phases of matter \cite{Wen,Haldane,Son,Kapustin}. Driven by these interconnected perspectives, we are predominantly interested in the study of the dynamical content of the Carlip-Gegenberg theory \cite{SCarlip}, known as  BCEA model, where a pair of Lorentz Lie algebra-valued 1-form gauge fields is coupled to 3D gravity in the first-order formalism. This coupling is implemented via the spin connection through a Horowitz-like topological term \cite{Horowitz}.

The Carlip-Gegenberg model has been subject to significant study. Analysis of its field equations shows that it can be interpreted as a teleparallel theory of gravity, again coupled to topological matter, in which the gravitational interaction is attributed to torsion or non-metricity rather than curvature. Notably, this type of coupling gives rise to exotic black hole solutions in which the roles of mass and angular momentum are reversed \cite{Mann}. Additionally, it has been established that the model admits a supersymmetric generalization \cite{Papadopolous}, can be extended to arbitrary dimensions \cite{Gegenberg}, and can be rewritten, up to surface terms, as a Chern-Simons theory \cite{Freidel2}. Despite these advances, the corresponding study of its Hamiltonian structure—including the full constraint structure, the gauge symmetry generators, and the resulting physical phase space—is still lacking. Importantly, in the context of condensed matter physics, recent studies reveal that although space-time-dependent effective actions can describe both the topological and dynamical aspects of topological matter, they often fail to capture all of their fundamental features, which are crucial for characterizing certain topological phases. In such cases, phase-space approaches  have emerged as more convenient and efficient theoretical tools; see \cite{Hayata,Palumbo} and references therein for a general discussion. Thus, providing such a  rigorous Hamiltonian analysis for the BCEA model at the non-perturbative level is one of the main motivations of the current work.

The Hamiltonian description provides a valuable and complementary perspective to its Lagrangian counterpart. This is especially critical for gravitational theories and their extensions, which are characterized by diffeomorphism invariance. As a gauge symmetry, diffeomorphism invariance leads to constraints that must be satisfied by the initial data. These constraints remove unphysical degrees of freedom—redundancies inherent to any covariant formulation—as is made explicit in the Hamiltonian framework. According to Dirac's prescription, the algebraic structure of these constraints—distinguishing between first-class (generating local symmetries) and second-class ones—is fundamental to the internal consistency of the theory and dictates the symplectic structure of the physical phase space \cite{Bergmann,Dirac,Henneaux,Rothe,Blagojevic}. Consequently, a complete Hamiltonian formulation that elucidates this full constraint structure serves as an indispensable foundation for two key objectives: first, achieving a deep understanding of the phase space and local symmetries with the prescription of the initial data on a Cauchy surface; and second, constructing a non-perturbative quantum theory. In this quantum context, first-class constraints become conditions that define the physical Hilbert space, Poisson-Dirac brackets are promoted to commutators for quantum observables, and the Hamiltonian constraint  governs the quantum dynamics \cite{Rovelli,Thiemann,Zachos}. Thus, it is highly desirable to investigate and classify, in a systematic manner, all constraints of the BCEA model via a fully non-perturbative Dirac Hamiltonian analysis \cite{Bergmann,Dirac}, as detailed below.

This paper is organized as follows. In Section \ref{The-Model}, we briefly present the Carlip-Gegenberg model, derive its field equations, and discuss its teleparallel interpretation. In Section \ref{Hamiltonian-formulation}, we cast the theory into ($2+1$) form and identify the initial phase space and primary constraints. We then implement the consistency algorithm, leading to the identification of the full set of secondary constraints. These constraints are subsequently classified into first- and second-class, enabling an explicit counting of the physical degrees of freedom without resorting to linearization. Furthermore, we construct the extended Hamiltonian, which generates the time evolution on the constraint surface. In Section \ref{Gauge-symmetries-of-the-action}, we employ the Castellani algorithm to build the gauge generators from the first-class constraints and derive the corresponding gauge transformations. We demonstrate how these transformations recover the diffeomorphism and local Poincaré symmetries of the theory, thereby connecting the constraint structure with the fundamental symmetries of the action. Section \ref{Dirac structure} is dedicated to the elimination of the second-class constraints by computing the Dirac bracket matrix, which defines the fundamental algebra for the canonical variables in the reduced phase space. Finally, in Section \ref{Conclusions}, we present our conclusions.

\section{The theory}
\label{The-Model}
Let us consider a pair of one-form fields coupled  with pure gravity in three dimensions, ruled by the following action  principle \cite{SCarlip}:
\begin{equation}
\mathrm{S}[B,C,E,A]= \int\limits_{\mathcal{M}}\mathrm{d}^{3}x\,\varepsilon^{\mu\nu\gamma}\left[\frac{1}{2}E_{I\mu}R_{\nu\gamma}^{I}(A)+B_{I \mu}\mathscr{D}\,_{\left[\nu\right.}C_{\left.\gamma\right]}{^{I}}\right],
\label{action}
\end{equation}
where $\mathcal{M}$ is the spacetime manifold we are integrating on, and $\varepsilon^{\mu\nu\gamma}$ is the Levi-Civita density determined  by the convention $\varepsilon^{012}=1$. The Greek indices $\mu,\nu,\cdots=0,1,2$ are tensor indices with respect to a coordinate basis determined by the coordinates $x^{\mu}$ on $\mathcal{M}$, while the Latin capital indices $I, J, K,\cdots=1,2,3$ are rigid indices in the vector representation of the Lorentz group $SO(2,1)$. Here, and henceforth, we use the following convention: $V_{\left[\mu\right.}W_{\left.\nu\right]}=V_{\mu}W_{\nu}-V_{\nu}W_{\mu}$. The dynamical variables are: a dreibein $E_{\mu}^{I}$ determining the spacetime metric $g_{\mu\nu}=E_{\mu}^{I}E_{\nu}^{J}\eta_{IJ}$, where $\eta_{IJ}$ is the 3D Minkowski metric with signature $(-\,+\,+)$; a spin connection $A_{\mu}^{I}$; and two matter fields $B_{\mu}^{I}$ and $C_{\mu}^{I}$. Furthermore, the curvature $R^{I}_{\mu\nu}$ associated with the connection $A_{\mu}^{I}$ is defined as
\begin{equation}
R^{I}_{\mu\nu}=\partial\:_{\left[\mu\right.}A_{\left.\nu\right]}{^{I}}+\epsilon^{I}{_{JK}}A^{J}_{\mu}A_{\nu}^{K},  
\end{equation}
while  the torsion  $T_{\mu\nu}^{I}$ of spacetime is defined as
\begin{equation}    T_{\mu\nu}^{I}=\mathscr{D}_{\left[\mu\right.}E_{\left.\nu\right]}^{I},
\end{equation}
where $\mathscr{D}$ is the covariant derivative with respect to the spin connection $A_{\mu}^{I}$ acting on index $I$ only;
\begin{eqnarray}
\mathscr{D}_{\mu}V^{I}&=&\partial_{\mu}V{^{I}}+\epsilon^{I}{_{JK}}A^{J}_{\mu}V^{K}\nonumber\\
&=&\left[\mathscr{D}_{\mu}\right]^{I}{_{J}}V^{J}.
\end{eqnarray}
For future use, we  have defined a differential operator as follows:
\begin{equation}
\left[\mathscr{D}_{\mu}\right]^{I}{_{J}}=\delta^{I}_{J}\partial_{\mu}+\epsilon^{I}{_{KJ}}A^{K}_{\mu},
\end{equation}
where $\partial$ is the fiducial derivative operator and $\epsilon^{IJK}$ is the completely antisymmetric $SO(2,1)$ tensor. Before we proceed to the Hamiltonian formulation of the model, we will analyze the equations of motion.

Varying the action (\ref{action}) with respect to the fields $E_{\mu}^{I}$, $A_{\mu}^{I}$, $B_{\mu}^{I}$, and $C_{\mu}^{I}$,  and asking it to be stationary, we arrive at the equations of motion
\begin{eqnarray}
{\mathbf \delta E}_{\mu}^{I}&:&\varepsilon_{\mu}{^{\nu\gamma}}R^{I}_{\nu\gamma}=0,\nonumber\\
{\mathbf \delta A}_{\mu}^{I}&:&\varepsilon_{\mu}{^{\nu\gamma}}\left[T_{\nu\gamma}^{I}+4\epsilon^{I}{_{JK}}B_{\nu}^{J}C_{\gamma}^{K}\right]=0,\nonumber\\
{\mathbf \delta B}_{\mu}^{I}&:&\varepsilon_{\mu}{^{\nu\gamma}}\mathscr{D}_{\nu}C^{I}_{\gamma}=0,\nonumber\\
{\mathbf \delta C}_{\mu}^{I}&:&\varepsilon_{\mu}{^{\nu\gamma}}\mathscr{D}_{\nu}B^{I}_{\gamma}=0.
\label{EofMot}
\end{eqnarray}

The first equation implies that $A_{\mu}^{I}$ is flat, while the second equation shows that it carries the torsion sourced by $B_{\mu}^{I}$ and $C_{\mu}^{I}$. This means that the spin connection $A_{\mu}^{I}$ is generally not compatible with the dreibein $E_{\mu}^{I}$. Relatedly, to solve the last two equations of motion for the matter field, they must be expressed in terms of a torsion-free connection.

To cure this apparent incompatibility, let us introduce the contortion tensor $K^{I}_{\mu}$ as the difference between the spin connection $A^{I}_{\mu}$ and a Riemannian (Levi-Civita) spin connection $\Omega_{\mu}^{I}$ that satisfies the following condition:
\begin{equation}
\left[\partial_{\left[\nu\right.}E_{\left.\gamma\right]}^{I}+\epsilon^{I}{_{JK}}\Omega_{\left[\nu\right.}^{J}E_{\left.\gamma\right]}^{K} \right]=0.\label{Torsion}
\end{equation}
Thus, we have
\begin{equation}
    K^{I}_{\mu}= A_{\mu}^{I}-\Omega_{\mu}^{I},\label{ConTorsion}
\end{equation}
which are in one-to-one correspondence with the usual torsion $T_{\mu\nu}^{I}$ through the following condition:
\begin{equation}
    \epsilon^{I}{_{JK}}K_{\left[\mu\right.}^{J}E_{\left.\nu\right]}^{K}=T^{I}_{\mu\nu}\label{ConditionConnction}.
\end{equation}
Substituting this into the second equation of motion in (\ref{EofMot}), we find the following algebraic condition:
\begin{equation}
   \epsilon^{I}{_{JK}}K_{\left[\mu\right.}^{J}E_{\left.\nu\right]}^{K}=-2\epsilon^{I}{_{JK}}B_{\left[\mu\right.}^{J}C_{\left.\nu\right]}^{K},\label{coupling}
\end{equation}
which, given the invertibility of the dreibein, allows us to uniquely determine the contorsion $K_{\mu}^{I}$ of the connection $A_{\mu}^{I}$ in terms of the two matter fields and the dreibein.  The result is 
\begin{eqnarray}
    K_{\mu}^{I}&=&\epsilon^{I}{_{J}}{^{K}}\left[\delta^{\nu}_{\mu}\delta^{L}_{K}+\frac{1}{2}E_{\mu}^{L}E^{\nu}_{K}\right]\epsilon^{J}_{MN}B_{\left[\nu\right.}^{M}C_{\left.\gamma\right]}^{N}E^{\gamma}_{L}\nonumber\\
    &=&K_{\mu}^{I}(E,B,C),\label{ContortionFinal}
\end{eqnarray}
where $E^{\mu}_{I}$ is the inverse matrix of $E^{I}_{\mu}$.  It is also possible to find relations between the curvature $R_{\mu\nu}^{I}(A)$ and the Riemannian ones $\widetilde{R}_{\mu\nu}^{I}(\Omega)$ associated with the Levi-Civita connection, and between the differential operator $\mathscr{D}(A)$ and the differential operator $\widetilde{\mathscr{D}}(\Omega)$ of the Levi-Civita connection. They read
\begin{eqnarray}
    R_{\mu\nu}^{I}(A)&=&\widetilde{R}_{\mu\nu}^{I}(\Omega)+\widetilde{\mathscr{D}}_{\left[\mu\right.}K^{I}_{\left.\nu\right]}+\epsilon^{I}{_{JK}}K^{J}_{\mu}K^{K}_{\nu},\\
    \left[\mathscr{D}_{\mu}\right]^{I}{_{J}}(A)&=&\left[\widetilde{\mathscr{D}}_{\mu}\right]^{I}{_{J}}(\Omega)+\epsilon^{I}{_{KJ}}K^{K}_{\mu}.
\end{eqnarray}
With this, we can rewrite the field equations (\ref{EofMot}) respectively as: 
\begin{eqnarray}
    \varepsilon^{\gamma\mu\nu}\left[\widetilde{R}_{\mu\nu}^{I}(\Omega)+\widetilde{\mathscr{D}}_{\left[\mu\right.}K^{I}_{\left.\nu\right]}+\epsilon^{I}{_{JK}}K^{J}_{\mu}K^{K}_{\nu}\right]&=&0,\nonumber\\
\varepsilon^{\gamma\mu\nu}\widetilde{\mathscr{D}}_{\left[\nu\right.}E_{\left.\gamma\right]}^{I}&=&0,\nonumber\\
\varepsilon^{\gamma\mu\nu}\left[\widetilde{\mathscr{D}}_{\mu}C^{I}_{\nu}+\epsilon^{I}{_{JK}}K^{J}_{\mu}C^{K}_{\nu}\right]&=&0,\nonumber\\
\varepsilon^{\gamma\mu\nu}\left[\widetilde{\mathscr{D}}_{\mu}B^{I}_{\nu}+\epsilon^{I}{_{JK}}K^{J}_{\mu}B^{K}_{\nu}\right]&=&0.\label{NewEoM}
\end{eqnarray}
At this stage, it is easy to see that the second equation can be recognized as the conventional torsion-free condition that ensures that the connection $\Omega_{\mu}^{I}$ is compatible with the dreibein $E_{\mu}^{I}$. As long as the dreibein is invertible, we can solve  for the spin connection $\Omega_{\mu}^{I}$, 
\begin{eqnarray}
    \Omega_{\mu}^{I}&=&-{\mathbf{E}}^{-1}\varepsilon^{\nu\rho\sigma}\left[E_{\nu}^{I}E_{\mu J}-\frac{1}{2}E_{\mu}^{I}E_{\nu J}\right]\partial_{\rho}E_{\sigma}^{J}\nonumber\\
    &=&\Omega_{\mu}^{I}(E),\quad {\mathbf{E}}=\text{det}\,E_{\mu}^{I}.\label{15}
\end{eqnarray}
Substituting (\ref{15}) and (\ref{ContortionFinal}) back into the expression (\ref{ConTorsion}), we finally find

\begin{equation}
    A_{\mu}^{I}=\epsilon^{I}{_{J}}{^{K}}\left[\delta^{\nu}_{\mu}\delta^{L}_{K}+\frac{1}{2}E_{\mu}^{L}E^{\nu}_{K}\right]\epsilon^{J}_{MN}B_{\left[\nu\right.}^{M}C_{\left.\gamma\right]}^{N}E^{\gamma}_{L}-{\mathbf{E}}^{-1}\varepsilon^{\nu\rho\sigma}\left[E_{\nu}^{I}E_{\mu J}-\frac{1}{2}E_{\mu}^{I}E_{\nu J}\right]\partial_{\rho}E_{\sigma}^{J}.\label{RelationConnections}
\end{equation}
This, in turn, implies that the spin connection $A_{\mu}^{I}$ highlights the interplay between geometry ($\Omega$) and matter-induced torsion ($K$).  Observe that the Levi-Civita connection $\Omega_{\mu}^{I}$ does not appear in the first set of equations of motion (\ref{EofMot}) because it is related to the dreibein, the matter fields, and the torsionful connection $A_{\mu}^{I}$ in a nontrivial manner, as shown in Eq. (\ref{RelationConnections}). Under these circumstances, there are two important cases to be drawn \cite{SCarlip, Mann}. In the first case, the model can be interpreted as a theory of 3D gravity with a dreibein $E_{\mu}^{I}$ and a torsionless non-flat spin connection $\Omega_{\mu}^{I}$ coupled with two topological matter fields, $B_{\mu}^{I}$ and $C_{\mu}^{I}$. In the second case, the model could be viewed as a teleparallel theory of gravity, with a dreibein $E_{\mu}^{I}$ and a torsionful flat connection $A_{\mu}^{I}$, again coupled to two matter fields $B_{\mu}^{I}$ and $C_{\mu}^{I}$. In both cases, the geometry is determined by the metric  $g_{\mu\nu}=\eta_{IJ}E^{I}_{\mu}E^{J}_{\nu}$. For the sake of concreteness,  we will focus on the second case—a connection that is flat but not necessarily torsion-free.

\section{Hamiltonian formulation}
\label{Hamiltonian-formulation}
\subsection{Constraints and their stability}
As a necessary first step toward constructing the Hamiltonian formulation for the model, not only do we assume that the spacetime $\mathcal{M}$ is globally hyperbolic such that it may be foliated as $\Re \times \Sigma$, with $\Sigma$ being a Cauchy’s surface without boundary and $\Re$ an evolution parameter, but also that simultaneous proper  ($2+1$) decompositions exist for the dreibein $E$, for the connection $A$, and for the matter fields $B$ and $C$. Then, by performing the $(2+1)$ decomposition of our fields, the action (\ref{action}) can be equivalently written in the canonical form $S=\int_{\Re} \mathrm{d}t L$, where the Lagrangian is  given by
\begin{equation}
L=\int\limits_{\Sigma} \mathrm{d}^{2}x\,\Big[\varepsilon^{0ab}\dot{A}^{I}_{a}E_{b I}+2\varepsilon^{0ab}\dot{C}^{I}_{a}B_{b I}+E_{0}^{I}\mathfrak{A}_{I}+A_{0}^{I}\mathfrak{B}_{I}+B_{0}^{I}\mathfrak{C}_{I}+C_{0}^{I}\aleph_{I}\Big]. \label{Lagrangian 2+1}
\end{equation}
Here $a,b,c,\cdots$ are space indices; the dot over a field variable denotes its time derivative, and the quantities $\mathfrak{A}_{I}$, $\mathfrak{B}_{I}$, $\mathfrak{C}_{I}$, and $\aleph_{I}$ are defined below: 
\begin{eqnarray}
\mathfrak{A}_{I}&=&\frac{1}{2}\varepsilon^{0ab}R_{abI},\nonumber\\
\mathfrak{B}_{I}&=&\varepsilon^{0ab}\left[\mathscr{D}_{a}E_{bI}+2\epsilon_{IJK}B_{a}^{J}C_{b}^{K}\right],\nonumber\\
\mathfrak{C}_{I}&=&2\varepsilon^{0ab}\mathscr{D}_{a}C_{bI},\nonumber\\
\aleph_{I}&=&2\varepsilon^{0ab}\mathscr{D}_{a}B_{bI}.\label{Exp4}
\end{eqnarray}
The corresponding spatial components of the curvature and the covariant derivative have the following structure:
\begin{eqnarray} 
R^{I}_{ab}&=&\partial_{a}A_{b}{^{I}}-\partial_{b}A_{a}{^{I}}+\epsilon^{I}{_{JK}}A^{J}_{a}A_{b}^{J},\\
\mathscr{D}_{a}V^{I}&=&\left[\mathscr{D}_{a}\right]^{I}{_{J}}V^{J}=\partial_{a}V{^{I}}+\epsilon^{I}{_{JK}}A^{J}_{a}V^{K}.
\end{eqnarray}

Furthermore, we consider the $72$-dimensional phase space $\Gamma$ spanned by $36$ pairs of conjugate variables $(q^{m}(x),p_{m}(x))$. Explicitly, they are
\begin{eqnarray}
q^{m}(x)&=&(E^{I}_{0},E^{I}_{a},A^{I}_{0},A^{I}_{a},B^{I}_{0},B^{I}_{a},C^{I}_{0},C^{I}_{a}),\nonumber\\
 p_{m}(x)&=&( \pi_{I}^{0},\pi_{I}^{a}, \Pi_{I}^{0},\Pi_{I}^{a},p_{I}^{0},p_{I}^{a},P_{I}^{0},P_{I}^{a}),\label{CanonicalVar}
\end{eqnarray}
where $p_{m}$ are the canonically conjugate momenta defined by computing  variations of the Lagrangian with respect to the time derivatives of the configuration space variables $q^{m}$, $p_{m}=\delta L/\delta \dot{q}^{m}$. Then
 \begin{eqnarray}
\Pi_{I}^{a}&=&\varepsilon^{0ab}E_{bI},\nonumber\\
P_{I}^{a}&=&2\varepsilon^{0ab}B_{bI},
\label{momentum}
\end{eqnarray}
and all other momenta equal to zero. As is well known, there exists a Hamiltonian function and a Poisson bracket structure on $\Gamma$ that give rise to the dynamical system (\ref{action}). The Poisson bracket  for two functionals of the fields and their momenta (\ref{CanonicalVar}), $F(q,p)$, and $G(q,p)$, is defined as 
\begin{equation}
    \{F(x),G(y)\}=\int \mathrm{d}^{2}z \left[\frac{\delta F(x)}{\delta q^{m}(z)}\frac{\delta G(y)}{\delta p_{m}(z)}-\frac{\delta F(x)}{\delta p_{m}(z)}\frac{\delta G(y)}{\delta q^{m}(z)}\right].\label{BraPoissonGen}
\end{equation}
This definition immediately provides the canonical Poisson brackets between the fields and their momenta, which follows from setting $F=q^{m}$ and $G=p_{m}$.  The (non-vanishing) canonical Poisson brackets are
\begin{eqnarray}
\{E_{I}^{\alpha}(x),\pi_{\beta}^{J}(y)\}&=&\delta^{\alpha}_{\beta}\delta^{J}_{I}\delta^{2}(x-y),\nonumber\\
\{A_{I}^{\alpha}(x),\Pi_{\beta}^{J}(y)\}&=&\delta^{\alpha}_{\beta}\delta^{J}_{I}\delta^{2}(x-y),\nonumber\\
\{B_{I}^{\alpha}(x),p_{\beta}^{J}(y)\}&=&\delta^{\alpha}_{\beta}\delta^{J}_{I}\delta^{2}(x-y),\nonumber\\
\{C_{I}^{\alpha}(x),P_{\beta}^{J}(y)\}&=&\delta^{\alpha}_{\beta}\delta^{J}_{I}\delta^{2}(x-y),\label{BraPoisson}
\end{eqnarray}
where $\delta^{2}(x-y)$ is a two-dimensional delta function. 

From Eq (\ref{momentum}), we see by inspection that none of the conjugate momenta $p_{m}$ depend on the time derivatives of the canonical coordinates ${q}^{m}$, and therefore all elements of the $36$-dimensional Hessian matrix , defined by $H_{mn}=\delta p_{m}/\delta\dot{q}^{n}$, are identically zero. This means that the rank of the Hessian is zero and, accordingly, the dimension of its kernel is $36$; therefore, we expect $36$ primary constraints. From momenta (\ref{momentum}) the following primary constraints emerge:
\begin{eqnarray}
\mathfrak{a}_{I}=\pi_{I}^{0}\approx0;&\quad&\mathfrak{a}_{I}^{a}=\pi_{I}^{a}\approx0\nonumber\\
\mathfrak{b}_{I}=\Pi_{I}^{0}\approx0;&&\mathfrak{b}_{I}^{a}=\Pi_{I}^{a}-\varepsilon^{0ab}E_{bI}\approx0\nonumber\\
\mathfrak{c}_{I}=p_{1}^{0}\approx0;&&\mathfrak{c}_{I}^{a}=p_{I}^{a}\approx0\nonumber\\
\mathfrak{n}_{I}=P_{1}^{0}\approx0;&&\mathfrak{n}_{I}^{a}=P_{I}^{a}-2\varepsilon^{0ab}B_{bI}\approx0,
\end{eqnarray}
Let us now denote them collectively as
\begin{equation}
\phi^{m}=\left(\mathfrak{a}_{I},\mathfrak{a}_{I}^{a},\mathfrak{b}_{I},\mathfrak{b}_{I}^{a},\mathfrak{c}_{I},\mathfrak{c}_{I}^{a},\mathfrak{n}_{I},\mathfrak{n}_{I}^{a}\right)\approx0.\label{primary}
\end{equation}
In this stage,  the dynamics of our system do not take place in the original phase space $\Gamma$, but rather on a submanifold $\Gamma_{\text{P}}$ of dimension $36$ determined by primary constraints $\phi^{m}$. We will refer to this submanifold $\Gamma_{P}$ as the primary constraint surface. Here and in what follows, the symbol ``$\approx$'' is used to designate ``on the constraint surface". For consistency of the theory, we must ensure that the primary constraint surface is conserved during its evolution. This task begins with the construction of the primary Hamiltonian that preserves the constraints (\ref{primary});  in fact, the primary Hamiltonian should be precisely the canonical Hamiltonian improved by a linear combination of primary constraints.

By performing  a Legendre transformation, we easily obtain the following canonical Hamiltonian:
\begin{equation}
H_{C}=-\int \mathrm{d}^{2}x\,\left[E_{0}^{I}\mathfrak{A}_{I}+A_{0}^{I}\mathfrak{B}_{I}+B_{0}^{I}\mathfrak{C}_{I}+C_{0}^{I}\aleph_{I}\right],
 \label{CanonicalHamiltonian}
\end{equation}
and accordingly, the  primary Hamiltonian reads
\begin{equation}
H_{P}=H_{C}+\int \mathrm{d}^{2}x\,\lambda_{m}\phi^{m},\label{HamPri}
\end{equation}
where $\lambda_{m}=\left(\lambda_{(1)}^{I},\lambda_{(2)a}^{I},\lambda_{(3)}^{I},\lambda_{(4)a}^{I},\lambda_{(5)}^{I},\lambda_{(6)a}^{I},\lambda_{(7)}^{I},\lambda_{(8)a}^{I}\right)$ is a set of 36 as-yet-undetermined multipliers associated with the primary constraints $\phi^{m}$. In that way, if all the constraints $\phi^{m}$ are to be preserved under the primary Hamiltonian $H_{P}$, it must be the case that \cite{Henneaux}:
\begin{equation}
\dot{\phi}^{m}=\{\phi^{m}(x),H_{P}(y)\}=\{\phi^{m}(x),H_{C}(y)\}+\int \mathrm{d}^{2}y\boxdot_{(xy)}^{mn}\lambda_{n}(y)\approx0,\label{Consistence}
\end{equation}
with  \, $\boxdot_{(xy)}^{mn}$\, being an anti-symmetric $(36\times36)$ matrix composed of Poisson brackets of all the primary constraints,  $\boxdot_{(xy)}^{mn}=\{\phi^{m}(x),\phi^{n}(y)\}$. As a consequence of (\ref{BraPoisson}),  we find  the non-vanishing  elements  of $\boxdot_{(xy)}^{mn}$ in Eq. (\ref{Consistence})
\begin{eqnarray}
\{\mathfrak{a}^{a}_{I}(x),\mathfrak{b}^{b}_{J}(y)\}&=&-\epsilon^{0ab}\eta_{IJ}\delta^{2}(x-y),\nonumber\\
\{\mathfrak{c}^{a}_{I}(x),\mathfrak{n}^{b}_{J}(y)\}&=&-2\epsilon^{0ab}\eta_{IJ}\delta^{2}(x-y).\label{AlgPrimary}
\end{eqnarray}
From (\ref{AlgPrimary}), it follows immediately that the kernel and rank of $\boxdot_{(xy)}^{mn}$ are equal to $12$ and $24$, respectively. Hence, from the consistency of the primary constraints (\ref{Consistence}), we expect to find the $12$ secondary constraints and determine the multipliers $24$. Concretely, the stability  of  $
\mathfrak{a}_{I}$, $\mathfrak{b}_{I}$, $\mathfrak{c}_{I}$, and $\mathfrak{n}_{I},$ leads to the following  12 secondary constraints, respectively:
\begin{eqnarray}
\{\mathfrak{a}_{I},H_{P}\}&=&\mathfrak{A}_{I}\approx0,\nonumber\\
\{\mathfrak{b}_{I},H_{P}\}&=&\mathfrak{B}_{I}\approx0,\nonumber\\
\{\mathfrak{c}_{I},H_{P}\}&=&\mathfrak{C}_{I}\approx0,\nonumber\\
\{\mathfrak{n}_{I},H_{P}\}&=&\aleph_{I}\approx0,
\end{eqnarray}
where the explicit forms of the secondary constraints $\mathfrak{A}_{I}$, $\mathfrak{B}_{I}$, $\mathfrak{C}_{I}$, $\aleph_{I}$ are defined in (\ref{Exp4}). We will denote them collectively by  
\begin{equation}
\psi_{m}=\left(\mathfrak{A}_{I},\mathfrak{B}_{I},\mathfrak{C}_{I},\aleph_{I}\right)\approx0.  \label{SecondaryConstraints}
\end{equation}
Additionally, the consistency conditions for the remaining primary constraints produce:
\begin{eqnarray}
\{\mathfrak{a}^{a}_{I},H_{P}\}&=&\varepsilon^{0ab}\left[\mathscr{D}_{b}A_{0I}-\lambda_{(4)bI}\right]\approx0,\nonumber\\
\{\mathfrak{b}^{a}_{I},H_{P}\}&=&\varepsilon^{0ab}\left[\mathscr{D}_{b}E_{0I}-\epsilon_{IJK}\left[A_{0}^{J}E_{b}^{K}-2B_{0}^{J}C_{b}^{K}-2C_{0}^{J}B_{b}^{K}\right]-\lambda_{(2)bI}\right]\approx0,\nonumber\\
\{\mathfrak{c}^{a}_{I},H_{P}\}&=&2\varepsilon^{0ab}\left[\mathscr{D}_{b}C_{0I}-\epsilon_{IJK}A^{J}_{0}C^{K}_{b}-\lambda_{(8)bI}\right]\approx0,\nonumber\\
\{\mathfrak{n}^{a}_{I},H_{P}\}&=&2\varepsilon^{0ab}\left[\mathscr{D}_{b}B_{0I}-\epsilon_{IJK}A^{J}_{0}B^{K}_{b}-\lambda_{(6)bI}\right]\approx0.
\end{eqnarray}
From the last four equations, we can conclude that
the multipliers accompanying the primary constraints, $\mathfrak{a}^{a}_{I}$, $\mathfrak{b}^{a}_{I}$, $\mathfrak{c}^{a}_{I}$, and $\mathfrak{n}^{a}_{I}$, can be algebraically determined in terms of the  canonical variables,
\begin{eqnarray}
\lambda_{(2)b}^{I}&\approx&\mathscr{D}_{b}E_{0}^{I}-\epsilon^{I}{_{JK}}\left[A_{0}^{J}E_{b}^{K}-2B_{0}^{J}C_{b}^{K}-2C_{0}^{J}B_{b}^{K}\right],\nonumber\\
\lambda_{(4)b}^{I}&\approx& \mathscr{D}_{b}A_{0}^{I},\nonumber\\
\lambda_{(6)b}^{I}&\approx&\mathscr{D}_{b}B_{0}^{I}-\epsilon^{I}{_{JK}}A^{J}_{0}B^{K}_{b},\nonumber\\
\lambda_{(8)b}^{I}&\approx&\mathscr{D}_{b}C_{0}^{I}-\epsilon^{I}{_{JK}}A^{J}_{0}C^{K}_{b}.\label{multipliers}
\end{eqnarray}
This means that the primary constraints $\mathfrak{a}^{a}_{I}$, $\mathfrak{b}^{a}_{I}$, $\mathfrak{c}^{a}_{I}$, and $\mathfrak{n}^{a}_{I}$ may be considered second-class. There are no further constraints. We will confirm this conclusion below.

Now, we summarize the whole set of constraints obtained in the above analysis for reference:
\begin{eqnarray}
36\quad\text{Primary constraints:}&&\quad\phi^{m}=(\mathfrak{a}_{I},\mathfrak{a}_{I}^{a},\mathfrak{b}_{I},\mathfrak{b}_{I}^{a},\mathfrak{c}_{I},\mathfrak{c}_{I}^{a},\mathfrak{n}_{I},\mathfrak{n}_{I}^{a})\approx0.\nonumber\\ 12\quad\text{Secondary constraints:}&&\quad\psi^{m}=(\mathfrak{A}_{I}, \mathfrak{B}_{I}, \mathfrak{C}_{I}, \aleph_{I})\approx0.\label{SetConstraints}
\end{eqnarray}
We will denote them collectively by $\gamma^{m}=\{\phi^{m},\psi^{m}\}$. This leaves us with a complete set of independent constraints\footnote{The condition of independence means that none of the constraints $\gamma^{m}$ can be represented as a linear combination of the remainder with coefficients that depend on $q^{m}$ and $p_{m}$, while completeness means that any function of $q^{m}$ and $p_{m}$ that vanishes on the constraint surface can be expressed linearly in terms of $\gamma^{m}$  with coefficients that depend on $q^{m}$ and $p_{m}$ \cite{Dirac}.}, which defines a subspace $\Gamma_{\text{Phys}}$, dubbed physical phase space, of the original phase space $\Gamma$. This means that our system remains bound to this subspace $\Gamma_{\text{Phys}}$ at all times and never leaves it. On the other hand, it is known that the secondary constraints can be modified or changed without altering $\Gamma_{\text{Phys}}$, that is, without affecting the dynamics on the constraint surface $\Gamma_{\text{Phys}}$. The transformation of secondary constraints which preserves their polynomial nature and keeps the physical phase space fixed is a contraction of constraints $\gamma^{m}$ with zero modes of the matrix of the Poisson brackets among all constraints $\gamma^{m}$ \cite{Henneaux}. 

Bearing this in mind, let us now define the $(48\times48)$ matrix, whose entries are the Poisson brackets of all the constraints with each other as $\boxplus_{(xy)}^{mn}=\{\gamma^{m}(x),\gamma^{n}(y)\}$. And let us write it  in block matrix form as,
    \begin{equation}
    \boxplus_{(xy)}^{mn}\,=\,
\bordermatrix{ \{\cdot\,,\cdot\}& \phi^{n}(y)&\psi^{n}(y) \cr
              \phi^{m}(x) & \boxdot_{(xy)}^{mn}& \boxtimes_{(xy)}^{mn}\cr
              \psi^{m}(x) & -\boxtimes_{(yx)}^{nm} & \boxminus_{(xy)}^{mn} \cr}.\label{Matrix}
\end{equation}
with
\begin{equation}
 \boxtimes_{(xy)}^{mn}=\{\phi^{m}(x),\psi^{n}(y)\},\quad\boxdot_{(xy)}^{mn}=\{\phi^{m}(x),\phi^{n}(y)\},\quad\boxminus_{(xy)}^{mn}=\{\psi^{m}(x),\psi^{n}(y)\}.  
\end{equation}
Using the Poisson brackets (\ref{BraPoissonGen}), we find  that the non-vanishing elements of  $\boxtimes_{(xy)}^{mn}$ take the following form:
\begin{eqnarray}
\{\mathfrak{A}^{I}(x),\mathfrak{b}_{J}^{a}(y)\}&=&-\varepsilon^{0ab}\left[\mathscr{D}^{\mathbf{x}}_{b}\right]^{I}{_{J}}\delta^{2}(x-y),\nonumber\\
\{\mathfrak{B}^{I}(x),\mathfrak{a}_{J}^{a}(y)\}&=&-\varepsilon^{0ab}\left[\mathscr{D}^{\mathbf{x}}_{b}\right]^{I}{_{J}}\delta^{2}(x-y),\nonumber\\
\{\mathfrak{B}^{I}(x),\mathfrak{b}_{J}^{a}(y)\}&=&\varepsilon^{0ab}\epsilon^{I}{_{JK}}E^{K}_{b}\delta^{2}(x-y),\nonumber\\
\{\mathfrak{B}^{I}(x),\mathfrak{c}_{J}^{a}(y)\}&=&2\varepsilon^{0ab}\epsilon^{I}{_{JK}}C^{K}_{b}\delta^{2}(x-y),\nonumber\\
\{\mathfrak{B}^{I}(x),\mathfrak{d}_{J}^{a}(y)\}&=&2\varepsilon^{0ab}\epsilon^{I}{_{JK}}B^{K}_{b}\delta^{2}(x-y),\nonumber\\
\{\mathfrak{C}^{I}(x),\mathfrak{b}_{J}^{a}(y)\}&=&2\varepsilon^{0ab}\epsilon^{I}{_{JK}}C^{K}_{b}\delta^{2}(x-y),\nonumber\\
\{\mathfrak{C}^{I}(x),\mathfrak{d}_{J}^{a}(y)\}&=&-2\varepsilon^{0ab}\left[\mathscr{D}^{\mathbf{x}}_{b}\right]^{I}{_{J}}\delta^{2}(x-y),\nonumber\\
\{\aleph^{I}(x),\mathfrak{b}_{J}^{a}(y)\}&=&2\varepsilon^{0ab}\epsilon^{I}{_{JK}}B^{K}_{b}\delta^{2}(x-y),\nonumber\\
\{\aleph^{I}(x),\mathfrak{c}_{J}^{a}(y)\}&=&-2\varepsilon^{0ab}\left[\mathscr{D}^{\mathbf{x}}_{b}\right]^{I}{_{J}}\delta^{2}(x-y).
\end{eqnarray}
Additionally, the non-zero elements of $\boxdot_{(xy)}^{mn}$ were defined in Eq. (\ref{AlgPrimary}). Finally, all the elements of $\boxminus_{(xy)}^{mn}$ turn out to be equal to zero. With these results at hand, it is easy to convince oneself that the matrix  $\boxplus_{(xy)}^{mn}$ is singular, and thus it has the following set of non-trivial zero-modes:
\begin{eqnarray}
V^{1}_{m}(x)&=&\left(0,\left[\mathscr{D}^{\mathbf{x}a}\right]_{IJ},0,0,0,0,0,0,\eta_{IJ},0,0,0\right),\nonumber\\
V^{2}_{m}(x)&=&\left(0,-\epsilon_{IJK}E^{aK},0,\left[\mathscr{D}^{\mathbf{x}a}\right]_{IJ},0,-\epsilon_{IJK}B^{aK},0,-\epsilon_{IJK}C^{aK},0,\eta_{IJ},0,0\right),\nonumber\\
V^{3}_{m}(x)&=&\left(0,-2\epsilon_{IJK}C^{aK},0,0,0,\left[\mathscr{D}^{\mathbf{x}a}\right]_{IJ},0,0,0,0,\eta_{IJ},0\right),\nonumber\\
V^{4}_{m}(x)&=&\left(0,-2\epsilon_{IJK}B^{aK},0,0,0,0,0,\left[\mathscr{D}^{\mathbf{x}a}\right]_{IJ},0,0,0,\eta_{IJ}\right).\label{Zero-Modes}
\end{eqnarray}
We can now contract each zero-mode $V_{m}^{1,2,3,4}$ with the whole set of constraints $\gamma^{m}$ to find the modified form of the secondary constraints.
\begin{eqnarray}
\int\mathrm{d}^{2}x V^{1}_{m}(x)\gamma^{m}(x)&=&\widetilde{\mathfrak{A}}_{I}=\mathfrak{A}_{I}+\mathscr{D}_{a}\mathfrak{a}^{a}_{I}\approx0,\nonumber\\
\int\mathrm{d}^{2}x V^{2}_{m}(x)\gamma^{m}(x)&=&\widetilde{\mathfrak{B}}_{I}=\mathfrak{B}_{I}+\mathscr{D}_{a}\mathfrak{b}^{a}_{I}+\epsilon_{I}{^{JK}}\left(E^{a}_{J}\mathfrak{a}_{aK}+B^{a}_{J}\mathfrak{c}_{aK}+C^{a}_{J}\mathfrak{n}_{aK}\right)\approx0,\nonumber\\
\int\mathrm{d}^{2}x V^{3}_{m}(x)\gamma^{m}(x)&=&\widetilde{\mathfrak{C}}_{I}=\mathfrak{C}_{I}+\mathscr{D}_{a}\mathfrak{c}^{a}_{I}+2\epsilon_{I}{^{JK}}C^{a}_{J}\mathfrak{a}_{aK}\approx0,\nonumber\\
\int\mathrm{d}^{2}x V^{4}_{m}(x)\gamma^{m}(x)&=&\widetilde{\aleph}_{I}=\aleph_{I}+\mathscr{D}_{a}\mathfrak{n}^{a}_{I}+2\epsilon_{I}{^{JK}}B^{a}_{J}\mathfrak{a}_{aK}\approx0.\label{ConstraintsStructureFinal}
\end{eqnarray}
Let us finally denote them by $\widetilde{\psi}^{m}=\left(\widetilde{\mathfrak{A}}_{I},\widetilde{\mathfrak{B}}_{I},\widetilde{\mathfrak{C}}_{I},\widetilde{\aleph}_{I}\right)$. The consistency of the physical space phase $\Gamma_{Phys}$ is ensured by the following Poisson algebra of the modified secondary constraints:
\begin{eqnarray}
\{\widetilde{\mathfrak{A}}^{I}(x),\widetilde{\mathfrak{B}}^{J}(y)\}&=&2\epsilon_{IJK}\widetilde{\mathfrak{A}}^{K}\delta^{2}(x-y),\nonumber\\
\{\widetilde{\mathfrak{B}}^{I}(x),\widetilde{\mathfrak{B}}^{J}(y)\}&=&2\epsilon_{IJK}\widetilde{\mathfrak{B}}^{K}\delta^{2}(x-y),\nonumber\\
\{\widetilde{\mathfrak{C}}^{I}(x),\widetilde{\mathfrak{B}}^{J}(y)\}&=&2\epsilon_{IJK}\widetilde{\mathfrak{C}}^{K}\delta^{2}(x-y),\nonumber\\
\{\widetilde{\aleph}^{I}(x),\widetilde{\mathfrak{B}}^{J}(y)\}&=&2\epsilon_{IJK}\widetilde{\aleph}^{K}\delta^{2}(x-y),\\
\{\widetilde{\mathfrak{C}}^{I}(x),\widetilde{\aleph}_{J}(y)\}&=&4\epsilon^{I}{_{JK}}\widetilde{\mathfrak{A}}^{K}\delta^{2}(x-y).
\label{Albebra1}
\end{eqnarray}
The remaining Poisson brackets between the modified secondary constraints are exactly  zero. Furthermore, the non-vanishing Poisson brackets between the primary and modified secondary constraints are given by
\begin{eqnarray}
\{\widetilde{\mathfrak{A}}^{I}(x),\mathfrak{b}_{J}^{a}(y)\}&=&\epsilon^{I}{_{JK}}\mathfrak{a}^{aK}\delta^{2}(x-y),\nonumber\\
\{\widetilde{\mathfrak{B}}^{I}(x),\mathfrak{a}_{J}^{a}(y)\}&=&\epsilon^{I}{_{JK}}\mathfrak{a}^{aK}\delta^{2}(x-y),\nonumber\\
\{\widetilde{\mathfrak{B}}^{I}(x),\mathfrak{b}_{J}^{a}(y)\}&=&\epsilon^{I}{_{JK}}\mathfrak{b}^{aK}\delta^{2}(x-y),\nonumber\\
\{\widetilde{\mathfrak{B}}^{I}(x),\mathfrak{c}_{J}^{a}(y)\}&=&\epsilon^{I}{_{JK}}\mathfrak{c}^{aK}\delta^{2}(x-y),\nonumber\\
\{\widetilde{\mathfrak{B}}^{I}(x),\mathfrak{n}_{J}^{a}(y)\}&=&\epsilon^{I}{_{JK}}\mathfrak{n}^{aK}\delta^{2}(x-y),\nonumber\\
\{\widetilde{\mathfrak{C}}^{I}(x),\mathfrak{b}_{J}^{a}(y)\}&=&\epsilon^{I}{_{JK}}\mathfrak{c}^{aK}\delta^{2}(x-y),\nonumber\\
\{\widetilde{\mathfrak{C}}^{I}(x),\mathfrak{n}_{J}^{a}(y)\}&=&2\epsilon^{I}{_{JK}}\mathfrak{a}^{aK}\delta^{2}(x-y),\nonumber\\
\{\widetilde{\aleph}^{I}(x),\mathfrak{b}_{J}^{a}(y)\}&=&\epsilon^{I}{_{JK}}\mathfrak{n}^{aK}\delta^{2}(x-y),\nonumber\\
\{\widetilde{\aleph}^{I}(x),\mathfrak{c}_{J}^{a}(y)\}&=&2\epsilon^{I}{_{JK}}\mathfrak{a}^{aK}\delta^{2}(x-y).\label{Albebra2}
\end{eqnarray}
The above constraints algebra also ensures that there are no more constraints, and in the absence of new constraints, the Dirac procedure  to obtain the true set of constraints has terminated. So, we have the complete constraint structure of the theory.

\subsection{Classification of constraints}
Now,  we are in the position of grouping all constraints into two classes, as indicated in refs.  \cite{ Dirac,Henneaux,Rothe}. Concretely, the first-class constraint set is the maximal set consisting of all constraints for which its Poisson bracket with all constraints vanishes weakly (i.e., they vanish on $\Gamma_{Phys}$). If a constraint is not first-class, it is automatically second-class. From these definitions and the constraint algebra (\ref{AlgPrimary}), (\ref{Albebra1}), and (\ref{Albebra2}), one readily infers that we have a set of $24$ first-class constraints given by 
 \begin{eqnarray}
       \mathfrak{a}_{I}\approx0,&&\mathfrak{b}_{I}\approx0,\quad \mathfrak{c}_{I}\approx0, \quad\mathfrak{n}_{I}\approx0,\nonumber\\
       \widetilde{\mathfrak{A}}_{I}\approx0,&& \widetilde{\mathfrak{B}}_{I}\approx0,\quad \widetilde{\mathfrak{C}}_{I}\approx0,\quad \widetilde{\aleph}_{I}\approx0.   \label{Fisrt-Class-Constraints}
 \end{eqnarray}
At the same time, we have  a set of $24$ second-class constraints given by
\begin{equation}
    \mathfrak{a}^{a}_{I}\approx0,\quad \mathfrak{b}^{a}_{I}\approx0,\quad \mathfrak{c}^{a}_{I}\approx0,\quad \mathfrak{n}^{a}_{I}\approx0. \label{Second-Class-Constraints}
\end{equation}
It is worth mentioning that the complete constraint algebra reveals an interplay between first- and second-class constraints that has not been previously reported in the literature. In fact, we demonstrate that the constraint structure identified in this work differs from those presented in \cite{SCarlip,Freidel2} 

\subsection{Physical degrees of freedom}

 The above classification allows us to carry out the counting of physical degrees of freedom in theory. On the one hand, according to Dirac's procedure for constraint Hamiltonian systems, each first-class  constraint(FCC) must remove one degree of freedom because it is a constraint, and it must remove a second degree of freedom because there is an arbitrary specifiable multiplier associated with the first-class constraints, while each second-class constraint (SCC) reduces the number of degrees of freedom by one. On the other hand, after the $(2+1)$ decomposition of the fields, the initial phase space is  $72$ dimensional. So, we get the following physical degrees of freedom count \cite{Henneaux,Rothe}
\begin{equation}
    \left[\text{dim}||\Gamma||-2\times\# \text{FCC}-\#\text{SCC}\right]
    =\left[72-2\times24-24\right]=0.
\end{equation}
Thus, we conclude that three-dimensional gravity theory, minimally coupled to topological matter fields, lacks physical degrees of freedom; that is, it defines a topological field theory.

\subsection{Extended Hamiltonian}
 We can now introduce the total Hamiltonian,
\begin{equation}
H_{T}=H_{P}+\int \mathrm{d}^{2}x\,\widetilde{\lambda}_{m}\widetilde{\psi}^{m},\label{TotalHam}    
\end{equation}
that contains all the constraints obtained in the above analysis, including primary and modified secondary constraints, and new arbitrary multipliers $\widetilde{\lambda}_{m}=(\widetilde{\lambda}_{(1)}^{I},\widetilde{\lambda}_{(2)}^{I},\widetilde{\lambda}_{(3)}^{I},\widetilde{\lambda}_{(4)}^{I})$. After plugging in the multipliers belonging to the second-class constraints, as given in Eqs. (\ref{multipliers}), into the total Hamiltonian $H_{T}$ (\ref{TotalHam}), we end up with
\begin{eqnarray}
    H_{T}&=&\int \mathrm{d}^{2}x\,\left[\lambda_{(1)}^{I}\mathfrak{a}_{I}+\lambda_{(2)}^{I}\mathfrak{b}_{I}+\lambda_{(3)}^{I}\mathfrak{c}_{I}+\lambda_{(4)}^{I}\mathfrak{n}_{I}\right.\nonumber\\ &&\left.+\left[\widetilde{\lambda}_{(1)}^{I}-E_{0}^{I}\right]\widetilde{\mathfrak{A}}_{I}+\left[\widetilde{\lambda}_{(2)}^{I}-A_{0}^{I}\right]\widetilde{\mathfrak{B}}_{I}+\left[\widetilde{\lambda}_{(3)}^{I}-B_{0}^{I}\right]\widetilde{\mathfrak{C}}_{I}+\left[\widetilde{\lambda}_{(4)}^{I}-C_{0}^{I}\right]\widetilde{\aleph}_{I}\right].\label{TotalHamiltonian}
\end{eqnarray}
Noting that each first-class constraint must have an associated indefinite multiplier, we redefine the multipliers $\widetilde{\lambda}_{m}$ in terms of the fields that did not appear with time derivatives in the Lagrangian (\ref{Lagrangian 2+1}) as  $\widetilde{\lambda}_{(1)}^{I}\rightarrow 2E_{0}^{I}$, $\widetilde{\lambda}_{(2)}^{I}\rightarrow 2A_{0}^{I}$ , $\widetilde{\lambda}_{(3)}^{I}\rightarrow 2B_{0}^{I}$, $\widetilde{\lambda}_{(4)}^{I}\rightarrow 2C_{0}^{I}$. For such redefinitions, the total Hamiltonian (\ref{TotalHamiltonian}) becomes  the  extended Hamiltonian,
\begin{equation}
H_{E}=\int \mathrm{d}^{2}x\,\left[\lambda_{(1)}^{I}\mathfrak{a}_{I}+\lambda_{(2)}^{I}\mathfrak{b}_{I}+\lambda_{(3)}^{I}\mathfrak{c}_{I}+\lambda_{(4)}^{I}\mathfrak{n}_{I}+E_{0}^{I}\widetilde{\mathfrak{A}}_{I}+A_{0}^{I}\widetilde{\mathfrak{B}}_{I}+B_{0}^{I}\widetilde{\mathfrak{C}}_{I}+C_{0}^{I}\widetilde{\aleph}_{I}\right].
\end{equation}
At this stage, the generalized Hamiltonian dynamics for our system is already described by this extended Hamiltonian. Below, we will find out the gauge transformations that leave the extended Hamiltonian unchanged.

\section{Gauge Symmetries}
\label{Gauge-symmetries-of-the-action}

\subsection{Gauge symmetry generator}
Now we can derive the generator of gauge transformations of the action (\ref{action}). To do this, let us first denote the whole set of first-class constraints as 
\begin{equation}
\Sigma^{m}=\left(\mathfrak{a}_{I},\mathfrak{b}_{I},\mathfrak{c}_{I},\mathfrak{n}_{I},\widetilde{\mathfrak{A}}_{I},\widetilde{\mathfrak{B}}_{I},\widetilde{\mathfrak{C}}_{I},\widetilde{\aleph}_{I}\right).\label{FistClass}
\end{equation}
Then, following Dirac’s hypothesis \cite{Dirac,Henneaux,Rothe}, the most general expression for the generator of gauge transformations can be constructed through a linear combination of all first-class constraints.  
\begin{equation}
    G=\int \mathrm{d}^{2}x\,\alpha_{m}\Sigma^{m},\label{FirstGenGauge}
\end{equation}
$\alpha_{m}$ are arbitrary gauge parameters. However, not all the parameters $\alpha_{m}$ will be independent. Using the Castellani algorithm \cite{Castellani}, we find that the  generating functional (\ref{FirstGenGauge}) of gauge transformations can be written, in terms of four (indexed) independent gauge parameters $\sigma^{I}$, $\omega^{I}$, $\rho^{I}$, and $\kappa^{I}$, as 
\begin{eqnarray}
G&=&\int \mathrm{d}^{2}x\,\left[\mathcal{S}(\sigma)+\mathcal{O}(\omega)+\mathcal{R}(\rho)+\mathcal{K}(\kappa)\right],\label{Generator}
\end{eqnarray}
where
\begin{eqnarray}
\mathcal{S}&=&-\dot{{\sigma}}^{I}\mathfrak{a}_{I}+\sigma^{I}\left[\widetilde{\mathfrak{A}}_{I}+\epsilon_{IJK}A^{J}_{0}\mathfrak{a}^{K}\right],\nonumber\\
\mathcal{O}&=&-\dot{\omega}^{I}\mathfrak{b}_{I}+\omega^{I}\left[\widetilde{\mathfrak{B}}_{I}+\epsilon_{IJK}\left(A^{J}_{0}\mathfrak{b}^{K}+E^{J}_{0}\mathfrak{a}^{K}+B^{J}_{0}\mathfrak{c}^{K}+C^{J}_{0}\mathfrak{n}^{K}\right)\right],\nonumber\\
\mathcal{R}&=&-\dot{\rho}^{I}\mathfrak{c}_{I}+\rho^{I}\left[\widetilde{\mathfrak{C}}_{I}+\epsilon_{IJK}\left(A^{J}_{0}\mathfrak{c}^{K}+2C^{J}_{0}\mathfrak{a}^{K}\right)\right],\nonumber\\
\mathcal{K}&=&-\dot{\kappa}^{I}\mathfrak{n}_{I}+\kappa^{I}\left[\widetilde{\aleph}_{I}+\epsilon_{IJK}\left(A^{J}_{0}\mathfrak{n}^{K}+2B^{J}_{0}\mathfrak{a}^{K}\right)\right].
\end{eqnarray}
The variation of canonical variables $q^{m}$ and $p_{m}$ under a gauge transformation `$\delta_{G}$' parametrized by (\ref{Generator}) is given, respectively, by
\begin{equation}
\delta_{G}q^{m}(x)=\{q^{m}(x),G\}\quad\&\quad\delta_{G}p_{m}(x)=\{p_{m}(x),G\}.\label{NewGenerators2}   
\end{equation}
As a consequence of Eqs. (\ref{Generator}) and (\ref{NewGenerators2}) we thus find the  Hamiltonian transformation laws of the theory,
\begin{eqnarray}
\delta_{G}E^{I}_{0}&=&-\mathscr{D}_{0}\sigma^{I}+\epsilon^{I}{_{JK}}\left[\omega^{J}E_{0}^{K}+2\rho^{J}C_{0}^{K}+2\kappa^{J}B_{0}^{K}\right],\nonumber\\
\delta_{G}E^{I}_{a}&=&-\mathscr{D}_{a}\sigma^{I}+\epsilon^{I}{_{JK}}\left[\omega^{J}E_{a}^{K}+2\rho^{J}C_{a}^{K}+2\kappa^{J}B_{a}^{K}\right],\nonumber\\
\delta_{G}\pi^{I}_{0}&=&\epsilon^{I}{_{JK}}\omega^{J}\pi^{K}_{0},\nonumber\\
\delta_{G}\pi^{I}_{a}&=&\epsilon^{I}{_{JK}}\omega^{J}\pi^{K}_{a},\nonumber\\
\delta_{G}A^{I}_{0}&=&-\mathscr{D}_{0}\omega^{I},\nonumber\\
\delta_{G}A^{I}_{a}&=&-\mathscr{D}_{a}\omega^{I},\nonumber\\
\delta_{G}\Pi^{I}_{0}&=&\epsilon^{I}{_{JK}}\left[\omega^{J}\Pi^{K}_{0}+\sigma^{J}\pi^{K}_{0}+\rho^{J}p^{K}_{0}+\kappa^{J}P^{K}_{0}\right],\nonumber\\
\delta_{G}\Pi^{I}_{a}&=&-\varepsilon_{0ab}\left[\mathscr{D}_{b}\sigma_{I}+\epsilon^{I}{_{JK}}2\rho^{J}C^{K}_{a}\right]+\epsilon^{I}{_{JK}}\left[\omega^{J}\Pi^{K}_{a}+\sigma^{J}\pi^{K}_{a}+\rho^{J}p^{K}_{a}+\kappa^{J}P^{K}_{a}\right],\nonumber\\
\delta_{G}B^{I}_{0}&=&-\mathscr{D}_{0}\rho^{I}+\epsilon^{I}{_{JK}}\omega^{J}B_{0}^{K},\nonumber\\
\delta_{G}B^{I}_{a}&=&-\mathscr{D}_{a}\rho^{I}+\epsilon^{I}{_{JK}}\omega^{J}B_{a}^{K},\nonumber\\
\delta_{G}p^{I}_{0}&=&\epsilon^{I}{_{JK}}\left[\omega^{J}p^{K}_{0}+2\kappa^{J}\pi^{K}_{0}\right],\nonumber\\
\delta_{G}p^{I}_{a}&=&\epsilon^{I}{_{JK}}\left[\omega^{J}p^{K}_{a}+2\kappa^{J}\pi^{K}_{a}\right],\nonumber\\
\delta_{G}C^{I}_{0}&=&-\mathscr{D}_{0}\kappa^{I}+\epsilon^{I}{_{JK}}\omega^{J}C_{0}^{K},\nonumber\\
\delta_{G}C^{I}_{a}&=&-\mathscr{D}_{a}\kappa^{I}+\epsilon^{I}{_{JK}}\omega^{J}C_{a}^{K},\nonumber\\
\delta_{G}P^{I}_{0}&=&\epsilon^{I}{_{JK}}\left[\omega^{J}P^{K}_{0}+2\rho^{J}\pi^{K}_{0}\right],\nonumber\\
\delta_{G}P^{I}_{a}&=&\epsilon^{I}{_{JK}}\left[\omega^{J}P^{K}_{a}+2\rho^{J}\pi^{K}_{a}\right]-2\varepsilon_{0ab}\mathscr{D}^{b}\rho_{I}.\label{HamiltonGuageSymm}
\end{eqnarray}
From these Hamiltonian gauge transformations (\ref{HamiltonGuageSymm}), we can easily derive the covariant form of the gauge transformation of the fields defining the action (\ref{action}). They are
\begin{eqnarray}
\delta_{G}E^{I}_{\mu}&=&-\mathscr{D}_{\mu}\sigma^{I}+\epsilon^{I}{_{JK}}\left[\omega^{J}E_{\mu}^{K}+2\rho^{J}C_{\mu}^{K}+2\kappa^{J}B_{\mu}^{K}\right],\nonumber\\
\delta_{G}A^{I}_{\mu}&=&-\mathscr{D}_{\mu}\omega^{I},\nonumber\\
\delta_{G}B^{I}_{\mu}&=&-\mathscr{D}_{\mu}\rho^{I}+\epsilon^{I}{_{JK}}\omega^{J}B_{\mu}^{K},\nonumber\\
\delta_{G}C^{I}_{\mu}&=&-\mathscr{D}_{\mu}\kappa^{I}+\epsilon^{I}{_{JK}}\omega^{J}C_{\mu}^{K}.\label{gaugefields}
\end{eqnarray}

\subsection{Diffeomorphism and Poincar\'e symmetries}
As a geometric theory describing the dynamics of spacetime, our theory cannot be understood without spacetime diffeomorphism (Diff) symmetry `$\delta_{\text{Diff}}$'. An easy way of seeing such a symmetry  is to redefine the gauge parameters as
\begin{equation}
\sigma^{I}=-E^{I}_{\mu}\xi^{\mu},\quad\rho^{I}=-B^{I}_{\mu}\xi^{\mu},\quad\kappa^{I}=-C^{I}_{\mu}\xi^{\mu},\quad\omega^{I}=-A^{I}_{\mu}\xi^{\mu},\label{re-label-Diff}
\end{equation}
where $\xi^{\mu}$ is an arbitrary three-vector. Such that the substitution of (\ref{re-label-Diff}) into (\ref{gaugefields}) yields on-shell spacetime diffeomorphism  symmetry `$\delta_{\text{Diff}}$' for the basic fields of the theory,
\begin{eqnarray}
\delta_{\text{Diff}}E^{I}_{\mu}&=&\mathcal{L}
_{\xi}E^{I}_{\mu}+\varepsilon_{\mu}{^{\nu\gamma}}\xi_{\nu}{\mathbf \delta A_{\gamma}^{ I}},\nonumber\\
\delta_{\text{Diff}}A^{I}_{\mu}&=&\mathcal{L}
_{\xi}A^{I}_{\mu}+\varepsilon_{\mu}{^{\nu\gamma}}\xi_{\nu}{\mathbf \delta E}_{\gamma}^{ I},\nonumber\\
\delta_{\text{Diff}}B^{I}_{\mu}&=&\mathcal{L}
_{\xi}B^{I}_{\mu}+\varepsilon_{\mu}{^{\nu\gamma}}\xi_{\nu}{\mathbf \delta C_{\gamma}^{ I}},\nonumber\\
\delta_{\text{Diff}}C^{I}_{\mu}&=&\mathcal{L}
_{\xi}C^{I}_{\mu}+\varepsilon_{\mu}{^{\nu\gamma}}\xi_{\nu}{\mathbf \delta B_{\gamma}^{ I}}.\label{Diffeomorphism}
\end{eqnarray}
Here $\mathcal{L} _{\xi}$ is the Lie derivative along the vector $\xi$. Note further that the  action (\ref{action}) describes a matter theory; therefore, it must naturally have the Poincaré transformations `$\delta_{\text{PGT}}$' as  symmetries. To recover the Poincaré symmetry, we will re-express the gauge parameters as follows 
\begin{equation}
\sigma^{I}=E^{I}_{\mu}\zeta^{\mu},\quad\rho^{I}=B^{I}_{\mu}\zeta^{\mu},\quad\kappa^{I}=C^{I}_{\mu}\zeta^{\mu},\quad\omega^{I}=\varpi^{I}+A^{I}_{\mu}\zeta^{\mu},\label{re-label}
\end{equation}
where $\zeta^{\mu}$ and $\varpi^{I}$ are related to local coordinate translations and local Lorentz rotations, respectively, and together constitute the 6 independent gauge parameters of Poincaré symmetries in three-dimensional spacetime. By substituting  (\ref{re-label}) into (\ref{gaugefields}), we finally find
\begin{eqnarray}
\delta_{\text{PGT}}E^{I}_{\mu}&=&-E^{I}_{\nu}\partial_{\mu}\zeta^{\nu}-\zeta^{\nu}\partial_{\nu}E^{I}_{\mu}-\epsilon^{I}{_{JK}}E^{J}_{\mu}\varpi^{K}-\varepsilon_{\mu}{^{\nu\gamma}}\zeta_{\nu}{\mathbf \delta  A_{\gamma}^{ I}},\nonumber\\
\delta_{\text{PGT}}A^{I}_{\mu}&=&-\partial_{\mu}\varpi^{I}-\epsilon^{I}{_{JK}}A^{J}_{\mu}\varpi^{I}-A^{I}_{\nu}\partial_{\mu}\zeta^{\nu}-\zeta^{\nu}\partial_{\nu}A^{I}_{\mu}-\varepsilon_{\mu}{^{\nu\gamma}}\zeta_{\nu}{\mathbf \delta  E}_{\gamma}^{ I},\nonumber\\
\delta_{\text{PGT}}B^{I}_{\mu}&=&-B^{I}_{\nu}\partial_{\mu}\zeta^{\nu}-\zeta^{\nu}\partial_{\nu}B^{I}_{\mu}-\epsilon^{I}{_{JK}}B^{J}_{\mu}\varpi^{K}-\varepsilon_{\mu}{^{\nu\gamma}}\zeta_{\nu}{\mathbf \delta  C_{\gamma}^{ I}},\nonumber\\
\delta_{\text{PGT}}C^{I}_{\mu}&=&-C^{I}_{\nu}\partial_{\mu}\zeta^{\nu}-\zeta^{\nu}\partial_{\nu}C^{I}_{\mu}-\epsilon^{I}{_{JK}}C^{J}_{\mu}\varpi^{K}-\varepsilon_{\mu}{^{\nu\gamma}}\zeta_{\nu}{\mathbf \delta  B_{\gamma}^{ I}}.\label{PGTSYM}
\end{eqnarray}
These transformations (\ref{PGTSYM}) comprise the Poincaré symmetries, modulo terms proportional to the equations of motion defined in (\ref{EofMot}). 

 \section{Dirac brackets}
 \label{Dirac structure}
Once the whole set of second-class constraints $\chi^{m}=\left(\mathfrak{a}^{a}_{I},\mathfrak{b}^{a}_{I},\mathfrak{c}^{a}_{I},\mathfrak{n}^{a}_{I}\right)$ has been identified, it can be eliminated from the theory by defining a new symplectic structure for the system, which is called the Dirac bracket. To this end, let us define the Dirac matrix $\Box^{mn}_{(xy)}$ whose entries are the Poisson brackets between these second-class constraints, i.e. $\Box^{mn}_{(xy)}=\{\chi^{m}(x),\chi^{n}(y)\}$:
\begin{equation}
    \Box_{(xy)}^{mn}=
\left(\begin{matrix}
\{\mathfrak{a}^{a}_{I},\mathfrak{a}^{b}_{J}\} & \{\mathfrak{a}^{a}_{I},\mathfrak{b}^{b}_{J}\} & \{\mathfrak{a}^{a}_{I},\mathfrak{a}^{c}_{J}\}&\{\mathfrak{a}^{a}_{I},\mathfrak{n}^{b}_{J}\}\\
\{\mathfrak{b}^{a}_{I},\mathfrak{a}^{b}_{J}\} & \{\mathfrak{b}^{a}_{I},\mathfrak{b}^{b}_{J}\} & \{\mathfrak{b}^{a}_{I},\mathfrak{c}^{b}_{J}\}&\{\mathfrak{b}^{a}_{I},\mathfrak{n}^{b}_{J}\}\\
\{\mathfrak{c}^{a}_{I},\mathfrak{a}^{b}_{J}\} & \{\mathfrak{c}^{a}_{I},\mathfrak{b}^{b}_{J}\} & \{\mathfrak{c}^{a}_{I},\mathfrak{c}^{b}_{J}\}&\{\mathfrak{c}^{a}_{I},\mathfrak{n}^{b}_{J}\}\\
\{\mathfrak{n}^{a}_{I},\mathfrak{a}^{b}_{J}\}&\{\mathfrak{n}^{a}_{I},\mathfrak{b}^{b}_{J}\}&\{\mathfrak{n}^{a}_{I},\mathfrak{c}^{b}_{J}\}&\{\mathfrak{n}^{a}_{I},\mathfrak{n}^{b}_{J}\}
\end{matrix}
\right).
\end{equation}
The explicit form of $ \Box_{(xy)}^{mn}$ reads:
\begin{equation}
    \Box_{(xy)}^{mn}=
\left(\begin{matrix}
0 & -1 & 0& 0\\
-1 & 0 & 0 & 0\\
0 & 0 & 0& -2\\
0& 0& -2& 0
\end{matrix}
\right)\varepsilon^{0ab}\eta_{IJ}\delta^{2}(x-y).
\end{equation}
This matrix has a non-vanishing determinant; therefore, it is invertible. Thus, the inverse of the matrix $\Box_{(xy)}^{mn}$ is now given by
\begin{equation}
    \left(\Box_{(yz)}^{mn}\right)^{-1}=
\left(\begin{matrix}
0 & -1 & 0& 0\\
-1 & 0 & 0 & 0\\
0 & 0 & 0& -\frac{1}{2}\\
0& 0& -\frac{1}{2}& 0
\end{matrix}
\right)\varepsilon_{0bd}\eta^{JK}\delta^{2}(y-z).
\end{equation}
The Dirac bracket of the theory for two dynamical entities $\mathcal{O}_{1}$ and $\mathcal{O}_{2}$ of the canonical variables is defined as
\begin{equation}
    \{\mathcal{O}_{1}(x),\mathcal{O}_{2}(y)\}_{D}=\{\mathcal{O}_{1}(x),\mathcal{O}_{2}(y)\}-\int \mathrm{d}^{2}z\int\mathrm{d}^{2}w\{\mathcal{O}_{1}(y),\chi^{m}(z)\}\left(\Box_{(zw)}^{mn}\right)^{-1}\{\chi^{n}(w),\mathcal{O}_{2}(y)\},
\end{equation}
where $\{\mathcal{O}_{1}(x),\mathcal{O}_{2}(y)\}$ is the Poisson bracket between the two functionals $\mathcal{O}_{1}$ and $\mathcal{O}_{2}$. From this definition, we find that the fundamental Dirac brackets between the dynamical variables parameterizing the reduced phase space are given by
\begin{eqnarray}
    \{E^{a}_{I}(x),A^{b}_{J}(y)\}_{D}&=&\varepsilon^{0ab}\eta_{IJ}\delta^{2}(x-y),\nonumber\\    \{B^{a}_{I}(x),C^{b}_{J}(y)\}_{D}&=&\frac{1}{2}\varepsilon^{0ab}\eta_{IJ}\delta^{2}(x-y).
\end{eqnarray}
It is well known that the Dirac brackets could be useful for studying physical observables as well as for performing the quantization of the theory.

\section{conclusions}
\label{Conclusions}

In this paper, we investigated the dynamical content of the coupling between a pair of $1$-form topological matter fields and first-order pure gravity in three dimensions \cite{SCarlip}.

Our analysis began with the action principle, from which we derived the field  equations (\ref{EofMot})  through independent variations. These equations involve a torsionful flat spin-connection $A$, which is generically incompatible with the dreibein $E$.  To address this incompatibility, we decomposed $A$ into a torsionless connection $\Omega$ and a contorsion $K$ (\ref{ConTorsion}). By combining Eqs. (\ref{EofMot}) and (\ref{Torsion}) into a simple algebraic equation (\ref{coupling}), we  explicitly solved for contorsion in terms of the dreibein and matter fields (\ref{ContortionFinal}). The torsion-free nature of $\Omega$ further enabled its explicit determination solely in terms of the dreibein (\ref{15}). Although the torsionless connection does not appear explicitly in the field equations (\ref{EofMot}), our analysis reveals that it is implicitly encoded through the non-trivial relation given in Eq. (\ref{RelationConnections}).

We then constructed the Hamiltonian formalism for this system using the Dirac algorithm. The system exhibits a highly constrained structure, as expected for gravity coupled to topological matter. Specifically, the system exhibits thirty-six primary and twelve secondary constraints, out of which only twenty-four are first-class ones (\ref{Fisrt-Class-Constraints}), while the remaining twenty-four are second-class (\ref{Second-Class-Constraints}). A key aspect of our analysis involves studying the matrices of brackets between the constraints. For instance, through the analysis of the structure of the zero-modes (\ref{Zero-Modes}) of the whole matrix (\ref{Matrix}) of the Poisson brackets of all constraints with each other, we rigorously derived the complete structure of secondary constraints (\ref{ConstraintsStructureFinal}). With this at hand, we classified all
the constraints into first- and second-class ones. We then used such first-class constraints to construct the generating functional (\ref{Generator}) that yields the gauge transformations of all phase-space variables (\ref{HamiltonGuageSymm}).  From these transformations, we explicitly extracted the covariant form of the gauge symmetry for the dynamical fields (\ref{gaugefields}). Furthermore, we successfully recovered the full diffeomorphism (\ref{Diffeomorphism}) and Poincar\'e (\ref{PGTSYM}) symmetries through an appropriate mapping of the gauge parameters. Importantly, the correct identification of all the constraints allowed us to confirm that the physical phase space has zero dimension per space point, confirming the absence of local degrees of freedom in the theory. This means that the matter fields do not introduce any additional local degree of freedom, and therefore the Carlip-Gegenberg model preserves the topological and solvability properties of $3$D gravity. Furthermore, we were able to construct the Dirac brackets for the BCEA system by using the inverse of the so-called Dirac matrix, of which the entries are the Poisson brackets among the second-class constraints. Our results provide a robust Hamiltonian complement to the findings of Refs. \cite{SCarlip,Freidel2}, reinforcing the consistency of the model and providing a framework that may be useful for studying the initial-value problem or generalizations to higher dimensions.

\section*{ACKNOWLEDGMENTS}
We acknowledge partial support from SNII-SECIHTI and DADIP-UNISON grants No. USO315009514.

\end{document}